\documentclass[a4paper,11pt]{article}

\usepackage[normalem]{ulem}
\usepackage{color}

\usepackage[blocks]{authblk}
\usepackage[utf8]{inputenc}

\usepackage{hyperref} 
\usepackage{url}
\usepackage{graphicx}
\usepackage{amssymb}
\usepackage{amsmath}

\usepackage{multirow}

\usepackage{graphics}
\usepackage{natbib}
\graphicspath{{./Figure}}

\usepackage{xcolor}
\usepackage{textcomp}
\usepackage{orcidlink}

\title{A SiPM-Based RICH Detector with Timing Capabilities for Isotope Identification}

\author[a,1]{M.~N.~Mazziotta\,\orcidlink{0000-0001-9325-4672}}
\author[a]{L.~Congedo\,\orcidlink{0000-0003-4536-4644}}
\author[a]{G.~De~Robertis\,\orcidlink{0000-0001-8261-6236},}
\author[a,b]{M.~Giliberti\,\orcidlink{0009-0007-2835-2963}\,}
\author[a]{F.~Licciulli\,\orcidlink{0000-0002-6955-0321},}
\author[a,b]{A.~Liguori\,\orcidlink{0009-0001-4240-6362}\,}
\author[a,b]{L.~Lorusso\,\orcidlink{0000-0002-2549-4401},}
\author[a,b]{N.~Nicassio\,\orcidlink{0000-0002-7839-2951},}
\author[a,b]{G.~ Panzarini\,\orcidlink{0000-0002-2586-1021},}
\author[a,b]{R.~Pillera\,\orcidlink{0000-0003-3808-963X}}

\affil[a]{Istituto Nazionale di Fisica Nucleare (INFN), Sezione di Bari, \\ via Orabona 4, I-70126 Bari, Italy}
\affil[b]{Dipartimento di Fisica dell'Universit\`a e del Politecnico di Bari, \\ via Amendola 173, I-70126 Bari, Italy}

\affil[1]{Corresponding author: mazziotta@ba.infn.it}

\date{}
\begin{document}

\maketitle

\begin{abstract}
In this work, we present a novel compact particle identification (PID) detector concept based on Silicon Photomultipliers (SiPMs) optimized to perform combined Ring-Imaging Cherenkov (RICH) and Time-of-Flight (TOF) measurements using a common photodetector layer. The system consists of a Cherenkov radiator layer separated from a photosensitive surface equipped with SiPMs by an expansion gap. A thin glass slab, acting as a second Cherenkov radiator, is coupled to the SiPMs to perform Cherenkov-based charged particle timing measurements. We assembled a small-scale prototype instrumented with various Hamamatsu SiPM array sensors with pixel pitches ranging from 2 to 3 mm and coupled with 1 mm thick fused silica window. The RICH radiator consisted of a 2 cm thick aerogel tile with a refractive index of 1.03 at 400 nm. The prototype was successfully tested in beam test campaigns at the CERN PS T10 beam line with pions and protons. We measured a single-hit angular resolution of about 4 mrad at the Cherenkov angle saturation value and a time resolution better than 50 ps RMS for charged particles with Z = 1. The present technology makes the proposed SiPM-based PID system particularly attractive for space applications due to the limited detector volumes available. In this work, we present beam test results obtained with the detector prototype and we discuss possible configurations optimized for the identification of ions in space applications.
\end{abstract}

\section{Introduction}
\label{sec:intro}

A key strategy for charged particle identification (PID) involves combining Time-of-Flight (TOF) measurements with data from Cherenkov radiation emitted by high-energy particles, typically using a Ring-Imaging Cherenkov (RICH) detector. Additionally, the identification of nuclei is achieved by exploiting the proportionality of the energy deposition to $Z^2$, where $Z$ is the charge in elementary charge units. Mass separation is performed through the simultaneous measurement of the particle rigidity (using a spectrometer) and velocity (using {TOF and Cherenkov detectors}). Historically, TOF and RICH measurements have been performed using independent detectors.

Recent advances in photon sensor technology suggest the possibility of designing a compact system that combines TOF and RICH detectors by sharing the same photodetector layer. Silicon Photomultipliers (SiPMs) are ideal sensors for such a system, thanks to their excellent performance in terms of time resolution, photodetection efficiency, magnetic field-insensitivity and low material budget (see, e.g., \cite{Gundacker:2020cnv}). This integrated PID RICH-TOF system significantly reduces the overall size of an instrument and the number of read-out channels compared to current detectors, such as those used in AMS-02 \cite{AMS:2021nhj} and HELIX \cite{Wakely:2023iaq}.

The AMS-02 TOF provides the primary trigger for the experiment and performs three critical measurements for PID: velocity, direction (downgoing/upgoing particles) and charge. It consists of four layers of plastic scintillator paddles (two above and two below the spectrometer) with a time resolution of about 160 ps for singly charged particles. The distance between the upper and lower layers is about 1.2 m. The AMS-02 TOF is equipped with specialized fine-mesh PMTs coupled with tilted light guides to successfully mitigate the impact of the spectrometer fringe magnetic field on the time resolution~\cite{Bindi:2010zzb}.
Similarly, the {HELIX TOF detector} still uses two 1.6 m long layers of plastic scintillator paddles read-out with SiPMs, with a total separation of 2.3 m. The top Hamamatsu S13360-6050VE SiPMs. The timing resolution achieved is about 170 ps for singly charged particles~\cite{HELIX:2023twg}.

The AMS-02 RICH is a proximity-focusing detector designed to measure velocity and charge of cosmic rays~\cite{Giovacchini:2023ixx}. It is equipped with two Cherenkov radiators: aerogel ($n \approx 1.05$) covering the large external area, enabling the measurement of high-velocity particles ($\beta > 0.95$); sodium fluoride (NaF) ($n \approx 1.33$) in the central region, with a higher refractive index, allowing for the measurement of lower-velocity particles. A lateral mirror surrounds the expansion volume to reflect outward photons to be detected by PMTs, in particular, the photons produced in the NaF radiator. However, the PMT matrix layer is ring-shaped, with a large central aperture to minimize the material in front of the electromagnetic calorimeter. The HELIX RICH detector also uses aerogel tiles for the radiator system, with a refractive index of $n \approx 1.15$. The photodetector layer---located just above the TOF lower paddles---consists of SiPM arrays with 75 $\mu$m cells and 6 $\times$ 6 mm$^2$ pixels~\cite{Jeon:2023pie}.

The prompt emission of Cherenkov radiation is ideal for achieving ultimate timing performance for a TOF detector. Charged particles above the Cherenkov threshold crossing a thin radiator slab with high refractive index, such as synthetic quartz (fused silica), can produce a fast signal~\cite{Credo:2004qgy}. A radiator a few millimeters thick with a proper refractive index can be optically coupled to a fast photodetector sensor, like a SiPM array, to provide highly precise signals~\cite{Mazziotta:2025zxj,Mazziotta:2025kdr,Nicassio:2023kux,Carnesecchi:2023dfq}. In this way, the charged particle generates a narrow cluster of hits close to its track (track hits) with a high signal amplitude. This system, when combined with a RICH detector that uses, for istance, an aerogel radiator, provides a valuable and compact layout for a PID detector operating in a wide momentum range. In addition, by requiring a time matching between aerogel Cherenkov photons and track hits, it is possible to improve pattern recognition by discarding possible uncorrelated SiPM dark-count hits. A crucial requirement for achieving the target time resolution is the optimization of both the radiator material and its thickness, along with the optical coupling to the SiPM surface. 

In this work, we report on the results of beam test campaigns performed using different prototype configurations.
We assembled a few small-scale prototypes instrumented with various Hamamatsu SiPM array sensors, with pitches ranging from 2 to 3 mm. These arrays were coupled with different window materials acting as a Cherenkov radiator, such as fused silica and MgF$_2$, with a thickness ranging from 1 to 2 mm. The optical coupling was obtained by using either epoxy resin or silicone resin. We also assembled a RICH detector in a proximity configuration using an aerogel radiator. These prototypes have been successfully tested at the CERN PS T10 beam line since 2022. 

Based on these experimental results, we then propose a possible application for a RICH-TOF system combined with a spectrometer to identify isotopes for either accelerator or space applications.

\section{Materials and Methods}
\label{sec:meth}

Since 2022, within the framework of the ALICE 3 upgrade planned for LHC Run 5~\cite{ALICE:2022wwr}, we have developed a compact and fast SiPM-based RICH detector for the future ALICE 3 PID system~\cite{Altamura:2025ajm,Mazziotta:2025zxj,Pillera:2025iqx,Mazziotta:2025kdr,Nicassio:2023kux}.

\begin{figure}[!h]
   \centering
    \includegraphics[scale=0.42]{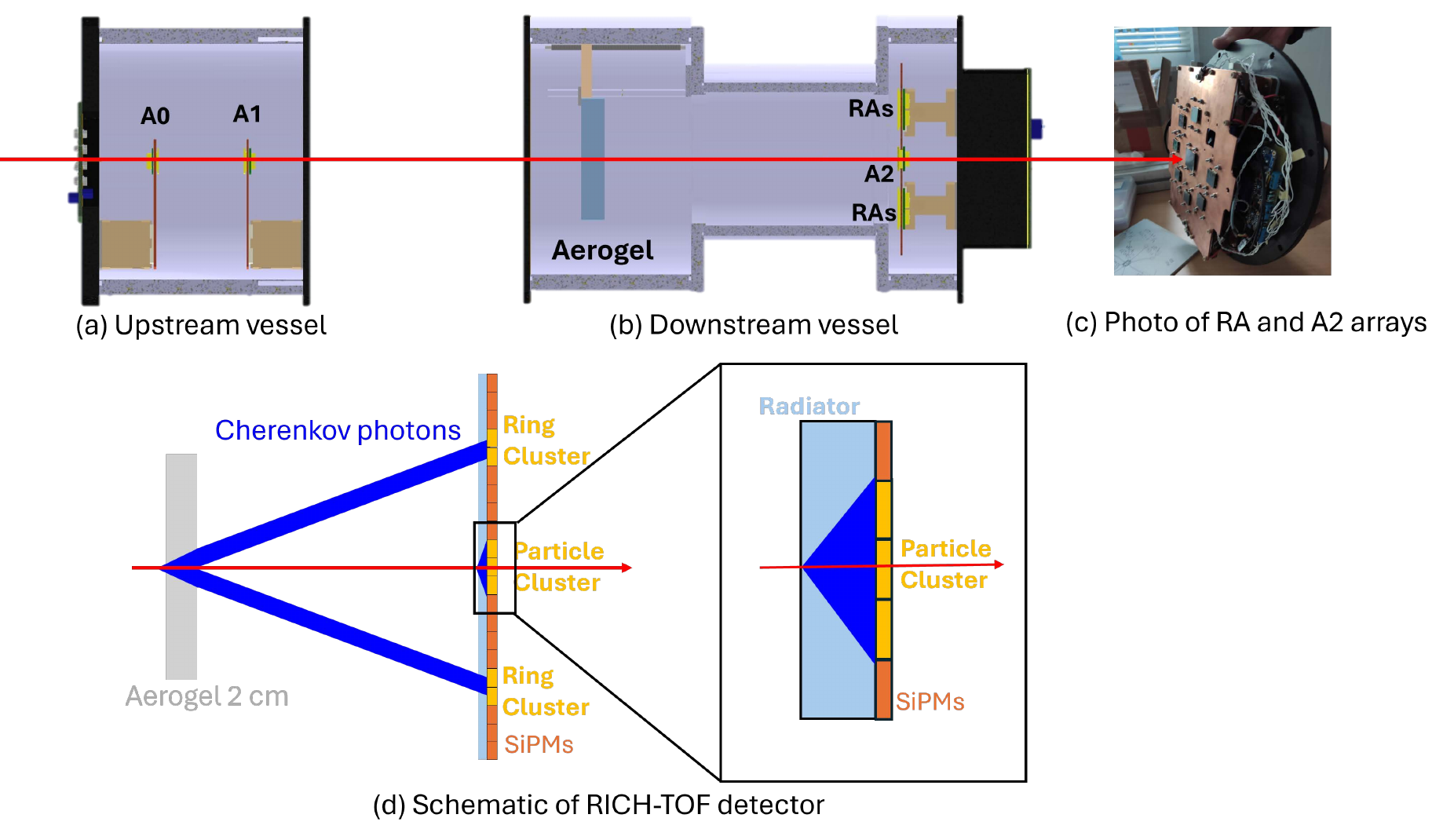}
    \caption{Prototype artist view showing the two-cylindrical vessel prototype and schematic of the proposed SiPM-based RICH-TOF system: (\textbf{a}) The upstream vessel houses two SiPM arrays (A0 and A1) at a distance of about 15 cm; (\textbf{b}) the downstream vessel houses the ring arrays (RAs) to collect the Cherenkov photons produced by the 2 cm thick aerogel radiator (located 23 cm upstream) and a central array (A2) to detect the hit produced by the charged particle in the thin window glued on the SiPM array, as shown in (\textbf{d}); (\textbf{c}) a photo of the photodetector layer is shown with the RAs and A2.}
    \label{fig:lay}
\end{figure}

For this R\&D project, we constructed a two-cylinder vessel prototype, as shown in the top right panel of Figure~\ref{fig:lay}. The upstream vessel (Figure~\ref{fig:lay}a) was equipped with two SiPM arrays (A0 and A1), which were separated by approximately 15 cm  and each featured a glued-in thin radiator window. The downstream vessel (Figure~\ref{fig:lay}b) housed an aerogel radiator and the photodetector layer (Figure~\ref{fig:lay}c). This setup was specifically designed to detect Cherenkov photons produced by charged particles in both the aerogel and the thin radiators (as illustrated in Figure~\ref{fig:lay}d). The aerogel radiator---a 2 cm thick, hydrophobic tile from {Aerogel Factory \& Co., Ltd.}
~\cite{ADACHI_Aerogel} with a refractive index {n}  = 1.03 at 400 nm wavelength~\cite{Altamura:2024xtz}---was positioned 23 cm upstream of the photodetector layer. This layer consisted of ring arrays (RAs), designed to detect photons from the aerogel, and an A2 array, which detected photons from the thin radiator (Figure~\ref{fig:lay}c). 
\vspace{-3pt}

\begin{figure}[!h]
    \centering    
    \includegraphics[scale=0.31]{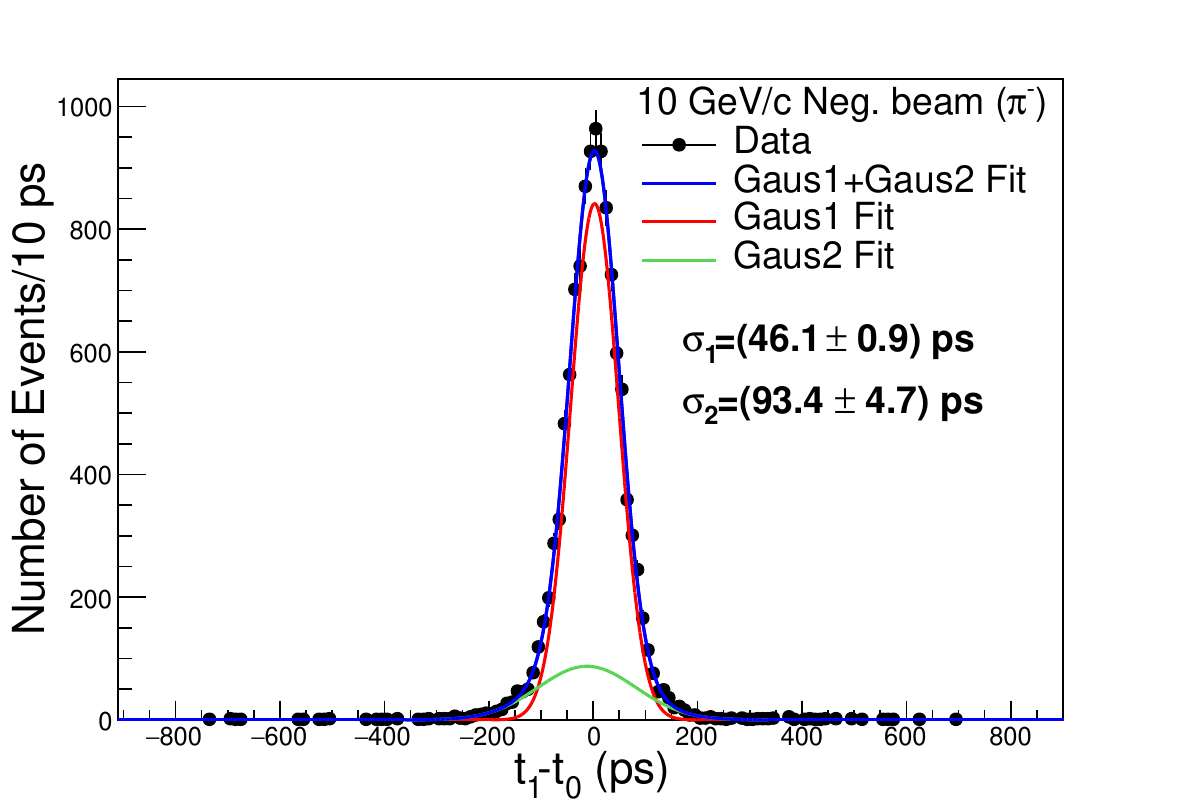}
    \includegraphics[scale=0.31]{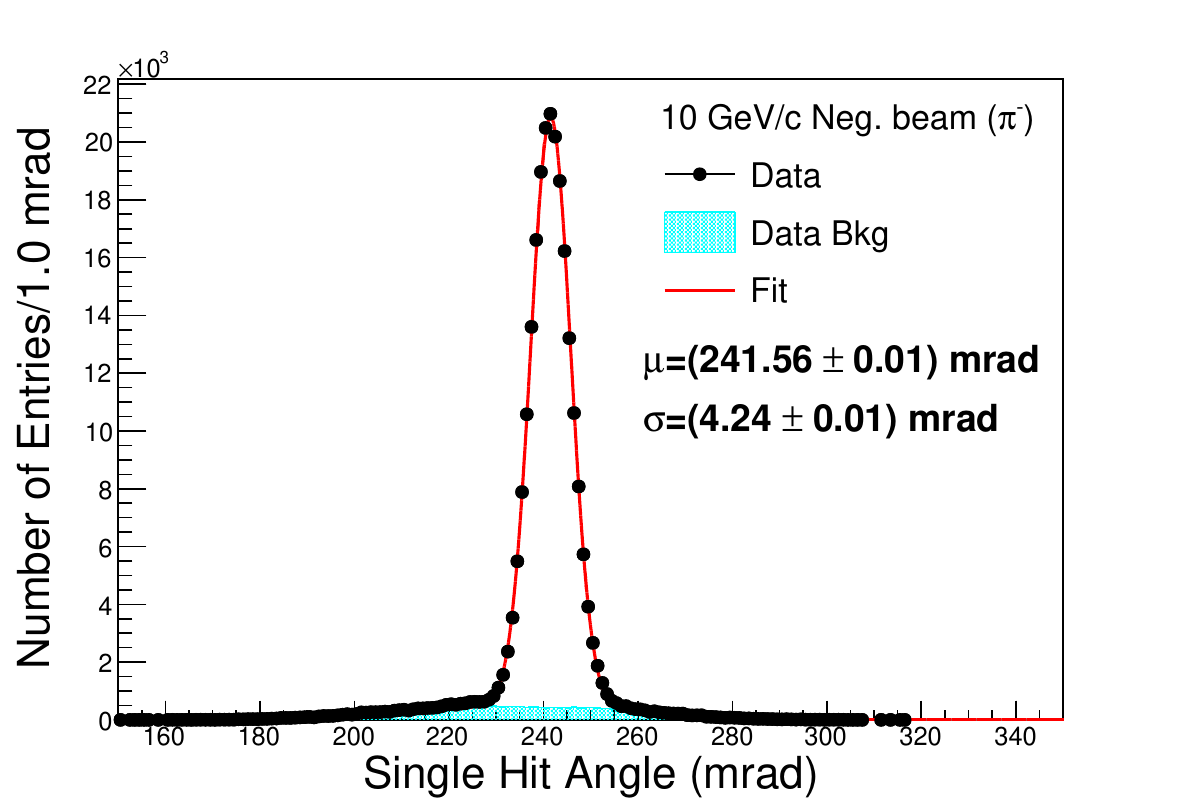}
    \caption{(\textbf{Left plot}): Distribution of the time differences by using the channels of S13361-2050AE-08 for A0 and S13361-3075AE-08 for the A2 arrays with 1 mm of fused silica glued on them. The histogram is fitted with two Gaussian distributions. The sigma of the core of the distribution (red line) is about 46 ps, while the sigma of the tail (green line) is about 93 ps. (\textbf{Right plot}): Cherenkov angle distribution measured with a 2 cm aerogel radiator and a proximity gap of about 23 cm by using a 2 mm SiPM pixel pitch. The single hit angular resolution is about 4.2 mrad in sigma unit~\cite{Mazziotta:2025zxj}.}
    \label{fig:timeangl}
\end{figure}

The SiPM arrays were mounted on custom printed circuit boards (SiPM carrier boards) attached to copper plates. These plates were cooled using a system of water chillers and Peltier cells. 
This setup enabled the SiPM arrays to operate at an average temperature of $-$5~$^\circ$C. This operating temperature is sufficient to mitigate uncorrelated hit noise resulting from the dark count rate of single photoelectrons, particularly for the RAs.
The vessels were flushed with nitrogen or argon to minimize humidity and thus maintain dew point values well below the minimum operating temperature inside the vessels.
The temperature of the copper plates was monitored using {TT4-10KC3-T125-M5-500 sensors}
~\cite{ntc10k} screwed on the plates, with data acquisition managed by a {{Raspberry Pi equipped with ADS1115 16-bit analog-to-digital converters (ADCs)}~\cite{ads1115}. Additionally, {SHT31-D}~\cite{sht31} sensors were used to monitor the humidity inside the vessels and in the surrounding room.

The prototypes were exposed to a mixed-particle beam (i.e., electrons, positrons, protons and pions) at the CERN T10-PS beam line, with momenta up to 10 GeV/c. Analog SiPM signals were routed from the vessel to external front-end electronics via 1 m long high-speed 50 Ohm multi-channel Samtec HLCD-20 (40 channels) cables~\cite{hlcd}. The data acquisition (DAQ) system consisted of a chain of front-end and readout electronics boards that included two front-end ASICs:

\begin{itemize}
    \item Petiroc 2A (developed by Weeroc) for charge and time measurements~\cite{Fleury:2014hfa,petiroc2a};
    \item Radioroc 2~\cite{Saleem:2023pwt} (also developed by Weeroc) as a front-end to amplify and then to discriminate the signals, followed by a picoTDC ASIC (developed by CERN)~\cite{Altruda:2023qoh}  to measure the arrival-time (ToA) and the time-over-threshold (ToT) of the analog SiPM signals.
\end{itemize}

A custom  front-end board (FEBs) housing the Petiroc 2A ASICs and an FPGA~\cite{Mazziotta:2022vow} was used for the DAQ, while the boards housing the Radioroc 2 and picoTDC ASICs were plugged to MOSAIC read-out boards~\cite{DeRobertis:2018vls}. Both FEBs were designed to read out the SiPM carrier boards, as previously discussed. This setup allowed us to use both DAQ systems to study the prototype performance.

For the A0, A1 and A2 arrays, we employed Hamamatsu Multi-Pixel Photon Counter (MPPC) SiPM devices, with $2\text{ mm}$ and $3\text{ mm}$ sizes and cell sizes of $50\ \mu \text{m}$ and $75\ \mu \text{m}$. We optically coupled $1\text{ mm}$ thick fused silica radiator windows (n $\approx 1.46$) to these arrays. Specifically, we used the following configurations:

\begin{enumerate}
    \item A0 and A1 with S13361-2050AE-08 arrays: $8 \times 8$ channels, $2\text{ mm}$ SiPM size, $50\ \mu \text{m}$ pixel size, and $2.2\text{ mm}$ pitch;
    \item A2 with S13361-3075AE-08 arrays: $8 \times 8$ channels, $3\text{ mm}$ SiPM size, $75\ \mu \text{m}$ pixel size, and $3.2\text{ mm}$ pitch.
\end{enumerate}

To test the effect of the radiator, we also used a bare S13361-3075AE-08 MPPC in the A0 position, which lacked a thin window radiator and instead we used the standard $100\ \mu \text{m}$ epoxy resin thickness. Furthermore, we measured the Cherenkov photon yield from a 3mm thick $\text{NaF}$ radiator (n $\approx 1.33$) by placing it $4\text{ mm}$ in front of a bare (but still with $100\ \mu \text{m}$ epoxy resin thick) S13361-30575AE MPPC in the A2 array. Finally, the arrays located in the ring consisted of a set of S13361-2050AE-08 SiPMs without any thin radiator window.

The beam test setup also included upstream and downstream X-Y fiber tracker modules, which were used for beam particle triggering and tracking~\cite{Mazziotta:2022vow,Pillera:2023hzp,Cerasole:2025tiz} (not shown in Figure~\ref{fig:lay}). With the setup discussed above (and shown in Figure~\ref{fig:lay}), we were able to measure the timing performance of the system by comparing the arrival times of the A0, A1 and A2 arrays. Additionally, using the RAs, we measured the RICH performance by detecting  Cherenkov photons produced in the aerogel, as will be discussed in Section~\ref{sec:res}.

\section{Results}
\label{sec:res}

The left side plot of Figure~\ref{fig:timeangl} shows the distribution of the time differences between the arrival times measured by the A0 and A2 arrays, by using a negative beam of 10 GeV/c that consists mainly of
pions~\cite{vanDijk:2025ggb}}. These arrays were assembled with S13361-2050AE-08 and S13361-3075AE-08, respectively, to which 1 mm of fused silica was glued. The measured times were first corrected for channel-by-channel time offsets and the time-walk effects, to account for signals with the same shape but different amplitudes crossing the discriminator threshold at different times. 
We used the Lookup Table (LUT) method to correct for time-walk. First, the average profile of the raw arrival time is determined as a function of the observed charge ($q$), where $q$ is measured as either photo-electron or Time-over-Threshold (ToT). Subsequently, the derived correction function, $f(q)$, is applied to each measurement: $t_{corr}=t_{raw} - f(q)$.

The measured time difference histogram is fitted with two Gaussian distributions to account for the core (ideal resolution) and the tail due to the deviation from the ideal core. The core of the time difference distribution, including SiPM and electronic time jitters, exhibits a $\sigma$ of about 46 ps. Assuming equal contributions from both arrays, this corresponds to a time resolution of approximately 35 ps for a single SiPM channel. The total number of photo-electrons measured with 1 mm of fused silica radiator windows was about 35--40 on average, which resulted in a particle detection efficiency (not corrected for the geometry and possible read-out electronic dead time) >99.5\% (see Ref.~\cite{Mazziotta:2025kdr} for more details). The timing performance of a similar system  without a thin radiator on A0 was also measured, which resulted in a reduced A0 efficiency and an overall time resolution worse than 200 ps (to be compared with the previous 46 ps). The worsening of the time resolution is a clear signature that the hit signals are due mainly to the Cherenkov photons produced in the thin radiator. The worsened resolution is due to the low number of observed hits and, consequently, a low detection efficiency. Those hits are still produced in the thin SiPM resin or due to a possible ionization loss in the SiPM depleted region.

The right side plot of Figure~\ref{fig:timeangl} shows the Cherenkov angle distribution measured with a 2 cm aerogel radiator and a proximity gap of $\approx$23 cm. The measurement was performed using 2 mm SiPM pixel pitch (S13361-2050AE-08 sensors) and a beam of negative 10 GeV/c mainly pions. We selected Cherenkov candidate events with arrival times within $\pm$5 ns with respect to the particle arrival times measured in the central array A2. The resulting single-photon Cherenkov angular resolution (in sigma units) was measured to be $(4.24 \pm 0.01)$ mrad. A significant level of background suppression (e.g., uncorrelated SiPM dark counts) was achieved by using the timing information and setting a coincidence window within a few nanoseconds~\cite{Altamura:2025ajm,Mazziotta:2025kdr,Mazziotta:2025zxj,Pillera:2025iqx}.

The Cherenkov photons emitted by 10 GeV/c pions in the 3 mm thick NaF radiator were measured with a HPK S13361-3075AE-08 64-channel SiPM array. The number of fired SiPM channels per event showed a Gaussian distribution with an average of 11.6 channels (see Ref.~\cite{Nicassio:2023kux} for more details).


\section{Perspective for Light Isotope Identification}

\begin{figure}[!h]
   \centering
    \includegraphics[scale=0.45]{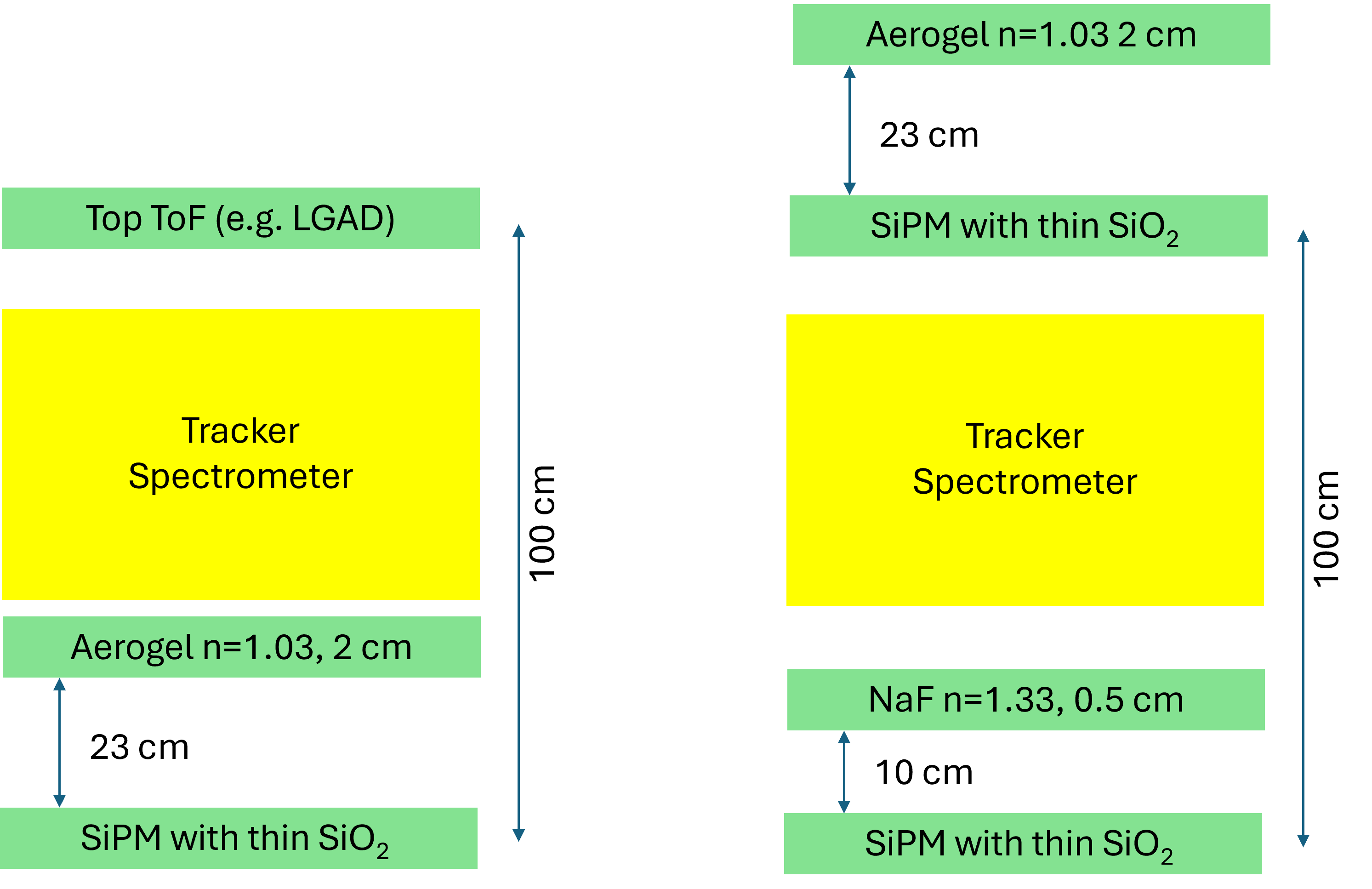}
    \caption{The figure shows two possible configurations for isotope identification. The left panel shows a proximity-focusing RICH with a dedicated top layer of TOF detector, made for instance of LGADs or SiPMs with a thin radiator window. The right panel illustrates a dual-RICH system with an aerogel and a NaF radiator, respectively.}
    \label{fig:NuclLay}
\end{figure}

Based on the beam test results discussed above, we can now outline potential applications for identifying nuclei, with a particular focus on the identification of Beryllium isotopes. Following the current layout of the AMS-02 instrument aboard the International Space Station (ISS)~\cite{Giovacchini:2023ixx}, Figure~\ref{fig:NuclLay} shows two possible configurations. The first configuration combines a proximity-focusing RICH detector with a dedicated top layer of a TOF detector, such as low-gain avalanche detectors (LGADs). In this case, the RICH photodetector layer is used to detect both the Cherenkov photons and to perform the time measurement to be combined with the LGAD one to have a compact instrument. The LGADs are based on a thin silicon sensor with internal gain and are capable of measuring with high precision the time with a resolution of $\mathcal{O}$(10~ps) with tracking capabilities~\cite{Sadrozinski:2013nja,Pellegrini:2014lki,Carnesecchi:2023jsu,Strazzi:2024tmn}. Of course, a stand-alone TOF layer could still consist of SiPM arrays with a thin radiator window optically coupled to the SiPMs.

\vspace{-3pt}

The second configuration uses a dual RICH detector, with both a low refractive index radiator (i.e., a 2 cm aerogel thick slab) and a high refractive index radiator (i.e., a 0.5 cm thick Sodium Fluoride, NaF, crystal). In this way, the two RICH detectors with their different radiator material allow us to extend the PID in a large momentum range and to provide a redundancy in the measurement of the Cherenkov angles. However, additional space is needed to accommodate the two RICH detectors. The RICH with the NaF radiator could be located downstream of the spectrometer to limit the material budget upstream. In both cases, the RICH photodetector layer consists of SiPMs with a 1 mm thick fused silica optically coupled to their top surfaces. Particle identification is achieved by also including a suitable  tracker--spectrometer in the layout. In the following, we will not discuss the tracker--spectrometer's performance, but we assume that it is sufficient to achieve particle mass identifications.

For the proposed layouts, we are assuming a total length of about one meter. The TOF time resolution is assumed to be 50 ps (i.e., 35 ps for each detection layer), while the RICH photodetector layer is assumed to use SiPMs with a read-out pixel size of 2 mm and a photodetection efficiency (PDE) of 40\% or more at 450 nm (e.g., SiPM with cell size of 50 $\mu$m). We are also assuming that the read-out electronics provide only the arrival time for each SiPM channel (hit). Also, for the current analysis, we are neglecting charge information, which could be obtained from methods such as the time-over-threshold or the slew-rate~\cite{Cossio:2024nju}. Charge measurements using ADC systems would provide a more powerful system.

To calculate the expected performance of such configurations, we used a fast simulation based on the beam test data discussed in Section~\ref{sec:res}. The beam test data, obtained with pions and protons, were scaled as a function of the momentum and by a factor $Z^2$, where $Z$ is the atomic number of the isotope. The scaling also accounts for the thickness of the Cherenkov radiator. Then, we calculated the number of detected Cherenkov photons, taking into account the SiPM PDE and that one hit for a 50 $\mu$m cell is allowed.

\begin{figure}[!h]
    \centering
    \includegraphics[scale=0.31]{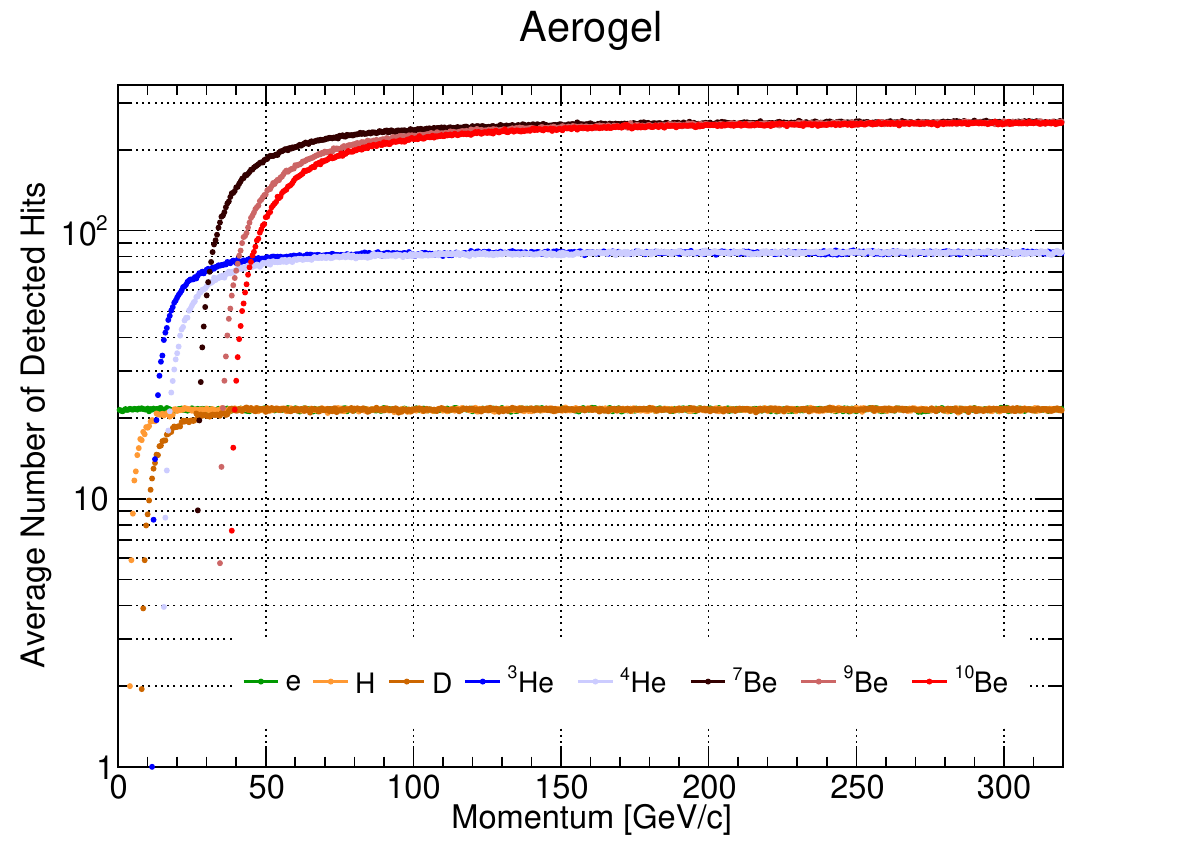}
    \includegraphics[scale=0.31]{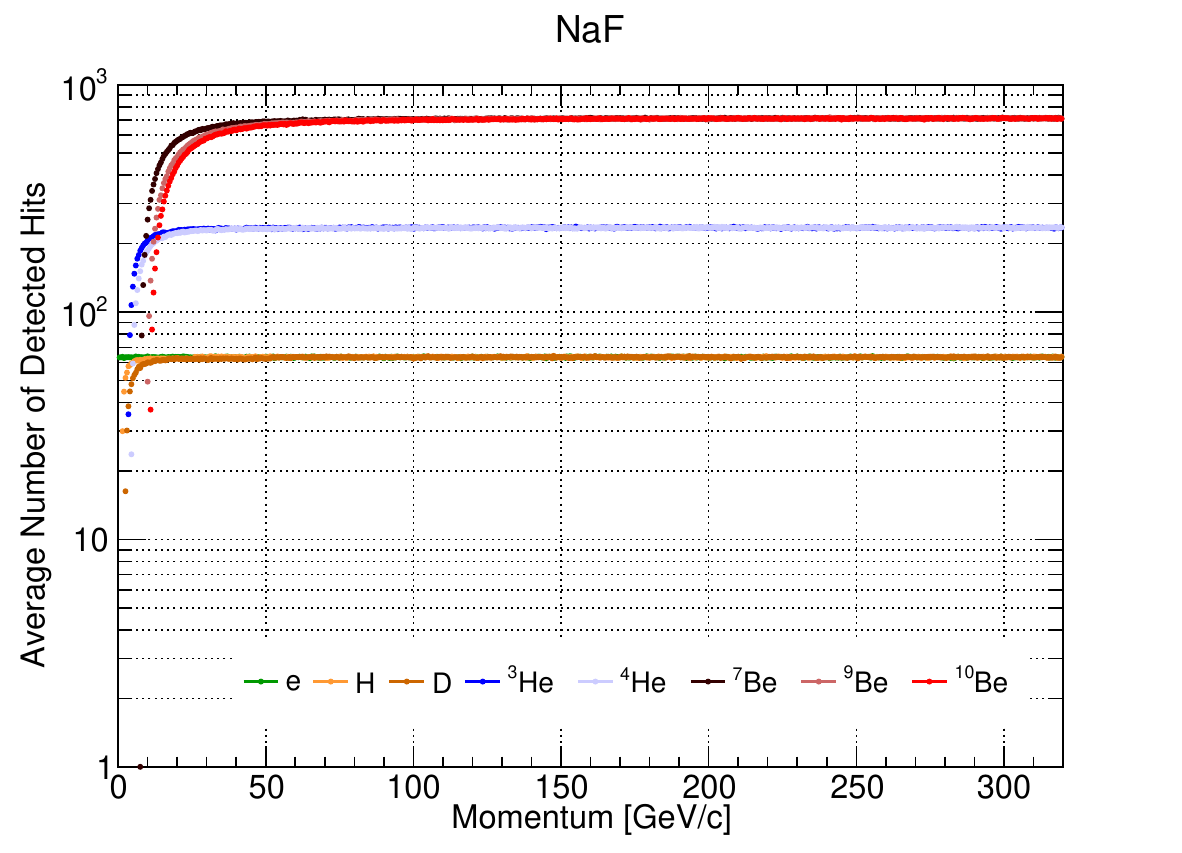}
    \caption{Expected average number of detected hits as a function of the momentum for electrons, protons and light nuclei for the aerogel- and NaF-based RICH configuration, respectively.}
    \label{fig:NpeNuc}
\end{figure}

\begin{figure}[!h]
   \centering
    \includegraphics[scale=0.31]{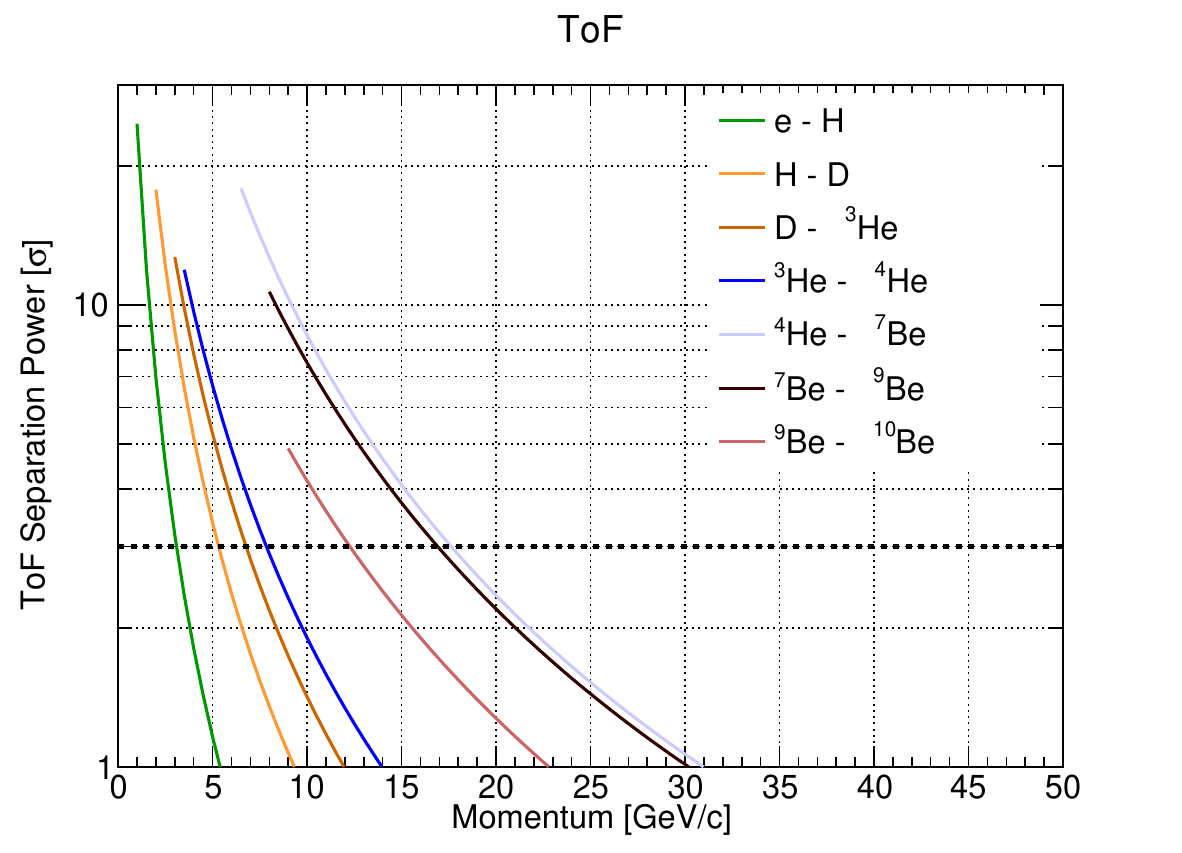}
    \includegraphics[scale=0.31]{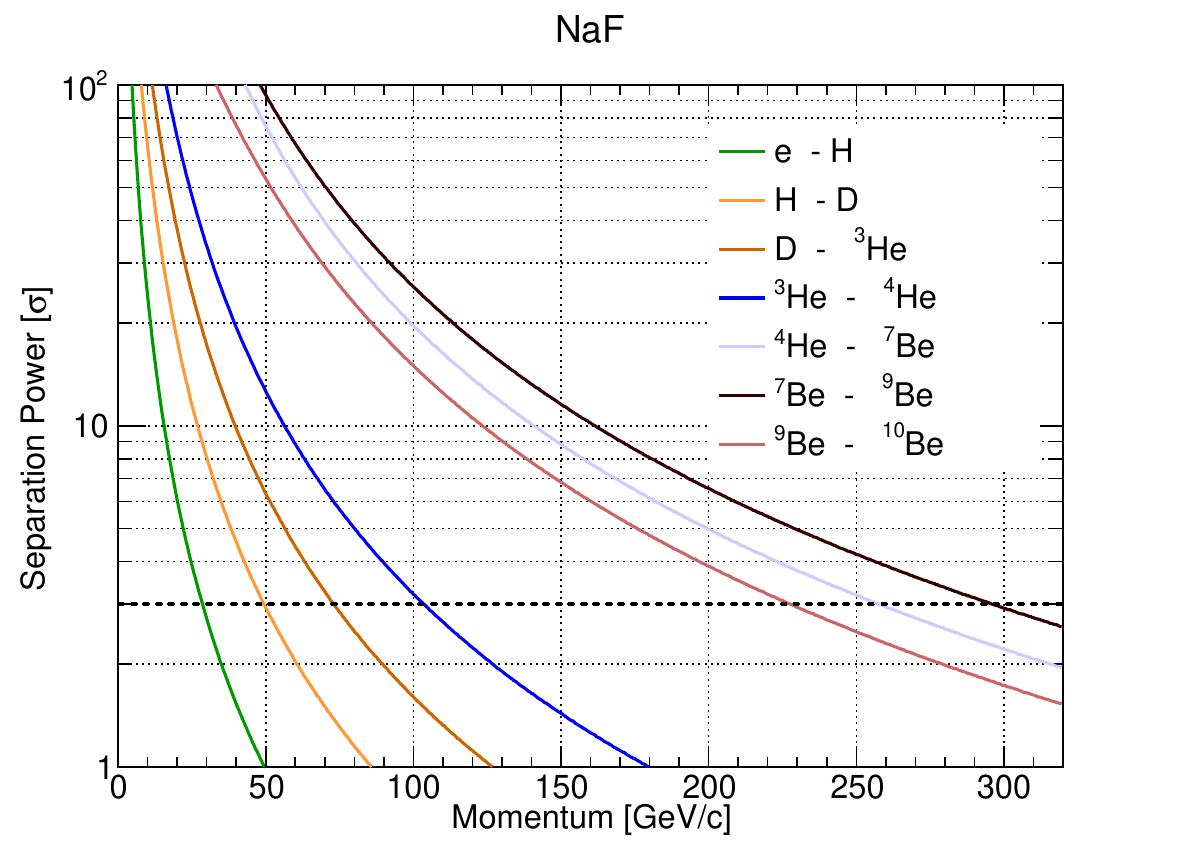}
    \includegraphics[scale=0.31]{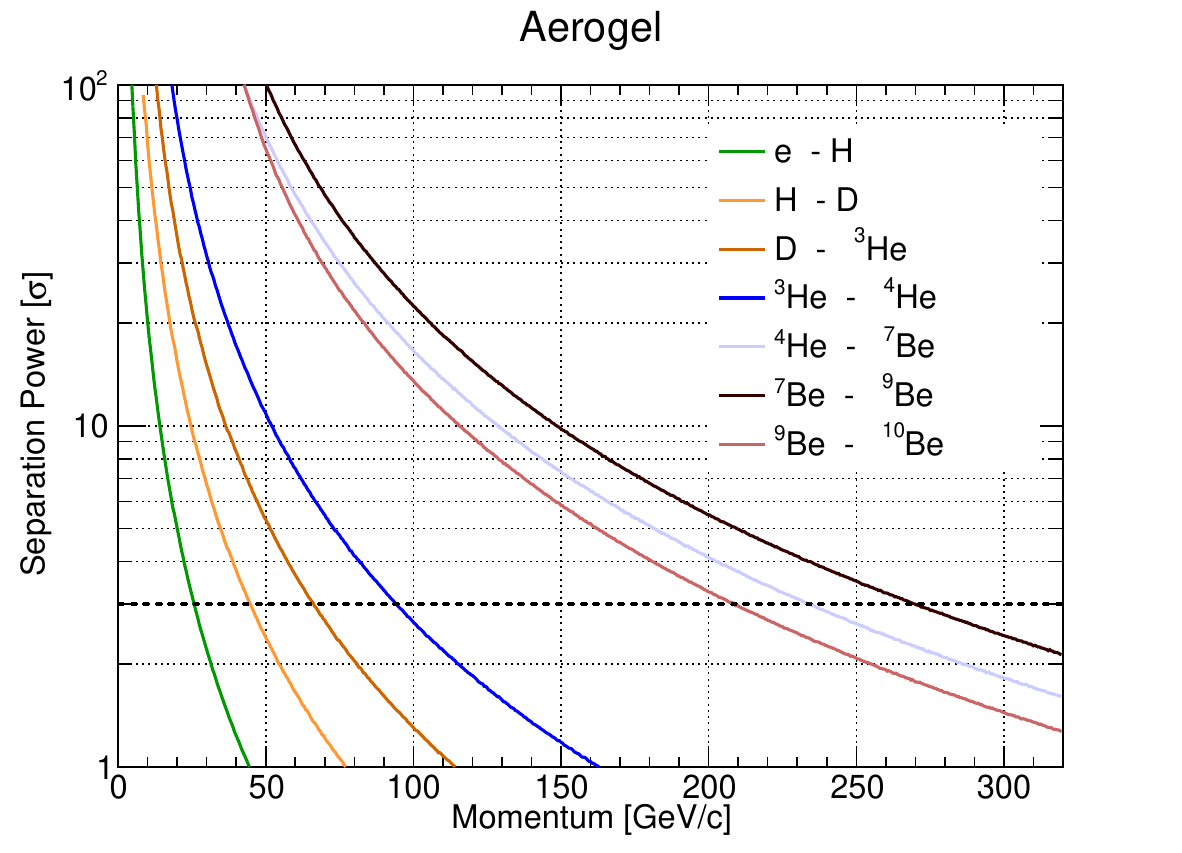}
    \caption{Separation power as a function of the momentum for electron-proton, proton-D and other various nuclei pairs for the TOF, NAF and aerogel RICH configuration (see text), respectively. }
    \label{fig:SepNuc}
\end{figure}

Figure~\ref{fig:NpeNuc} shows the expected number of hits for the aerogel-based and for the NaF-based RICHs as a function of the momentum for electrons, protons and several light isotopes. This corresponds to the configuration as shown in Figure~\ref{fig:NuclLay}. Based on the detected hits, Figure~\ref{fig:SepNuc} shows the expected separation power for the aerogel-based RICH and TOF detector as a function of the momentum for electron--proton (H), proton--deuteron (D), and several light nuclei species pairs. The maximum momentum value for the 3 $\sigma$ separation depends on the particle pair and system. For instance, in the case of $^4$He-$7$Be, it is about 18 GeV/c, 270 GeV/c and 300 GeV/c for the TOF, aerogel-RICH and NaF-RICH, respectively.

The identification of light nuclei isotopes like $^7$Be, $^9$Be and $^{10}$Be with the RICH and TOF system can be achieved by analyzing the mass {resolution  $\frac{dM}{M}$:

\begin{equation}
    \frac{dM}{M} = \sqrt{ \left( \gamma^2 \frac{d\beta}{\beta} \right)^2 + \left( \frac{d\mathcal{R}}{\mathcal{R}} \right)^2} \xrightarrow{d\mathcal{R}/\mathcal{R} \to 0} \gamma^2 \frac{d\beta}{\beta}
\end{equation}
where $\beta = \frac{|\vec{p}|}{E}$, $\gamma = \frac{1}{\sqrt{1-\beta^2}}$, $\mathcal{R}=\frac{|\vec{p}|}{Z}$ is the particle rigidity with momentum $\vec{p}$. In the following, we are neglecting the spectrometer term to the resolution, i.e., we are assuming $\frac{d\mathcal{R}}{\mathcal{R}}=0$.

Figure~\ref{fig:dM_M}a shows the mass separation for beryllium isotopes as a function of the momentum for the TOF, aerogel-RICH and NaF-RICH systems. In the considered momentum range, the mass separation is always below 20\%. Figures~\ref{fig:dM_M}b--d show the expected mass number distributions, again for the beryllium isotopes at specific momentum values for the TOF, NaF-RICH and aerogel-RICH configurations. The distributions are shown assuming an equal abundance of the isotopes. Indeed, the abundance of $^{10}$Be is estimated to be $\approx$20\% (energy-dependent) relative to that of $^9$Be, since its lifetime of approximately $\approx$1.4 Myr is comparable to the cosmic-ray propagation time in the Galaxy~\cite{delaTorreLuque:2022vhm}.

\begin{figure}[!h]
    \includegraphics[scale=0.34]{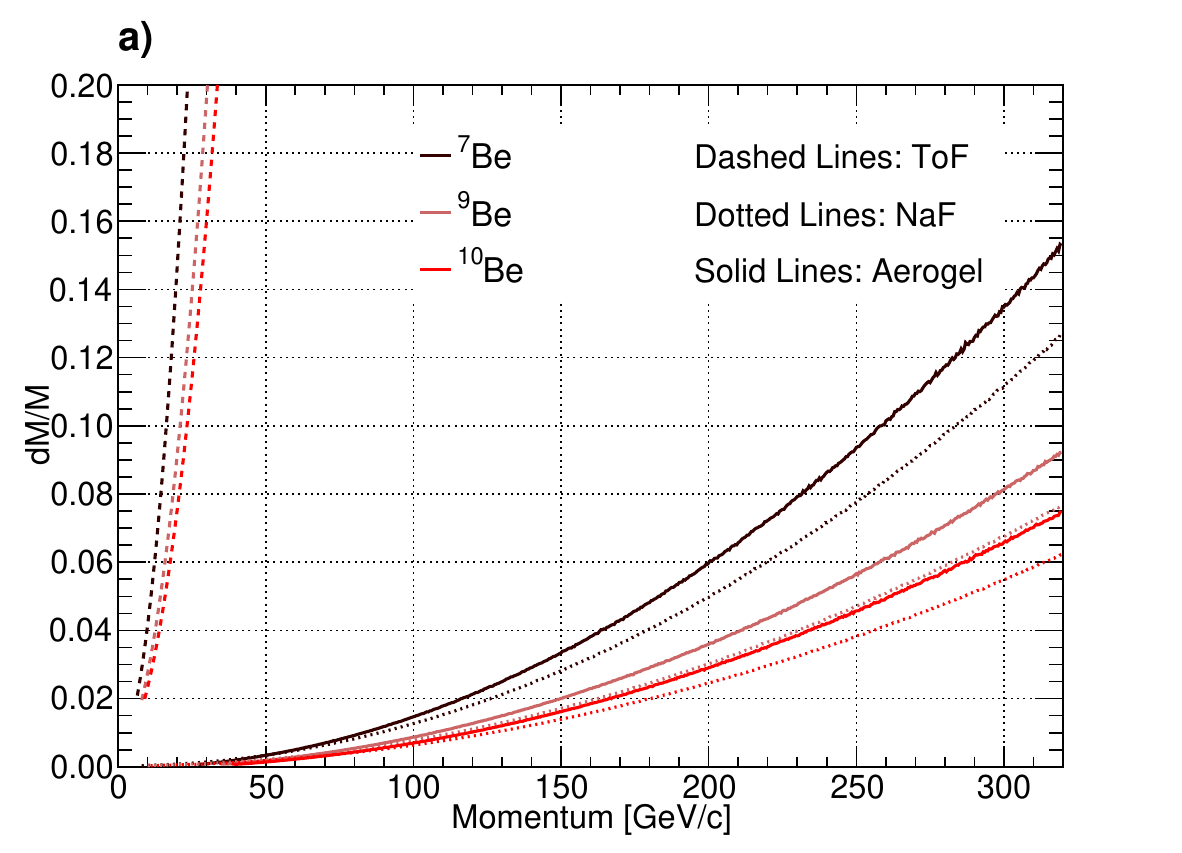}
    \includegraphics[scale=0.24]{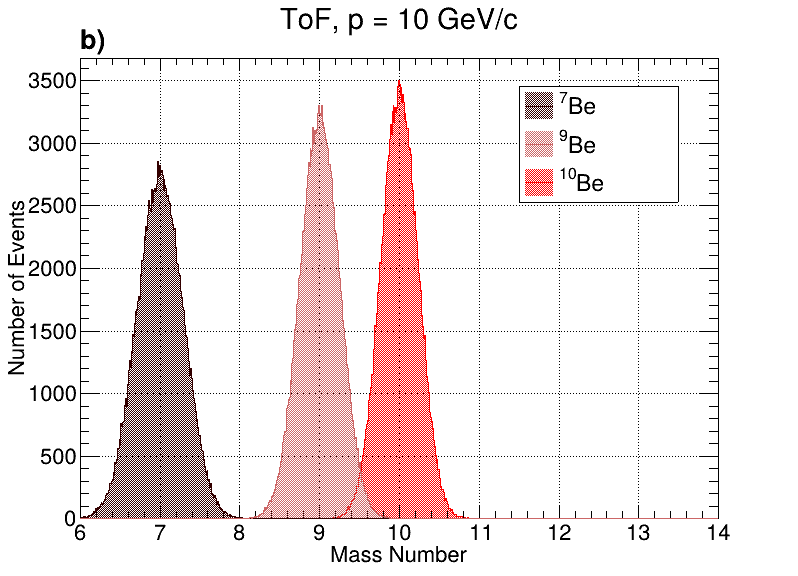}
    \includegraphics[scale=0.24]{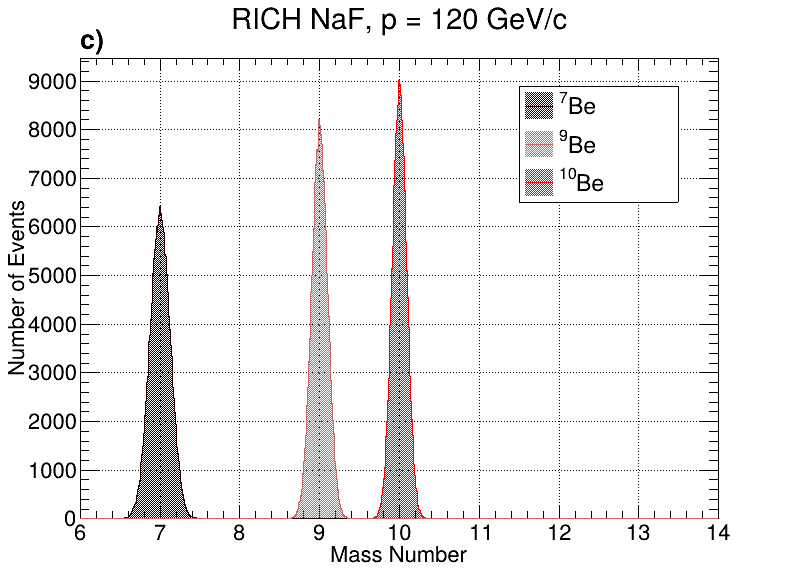}
    \includegraphics[scale=0.24]{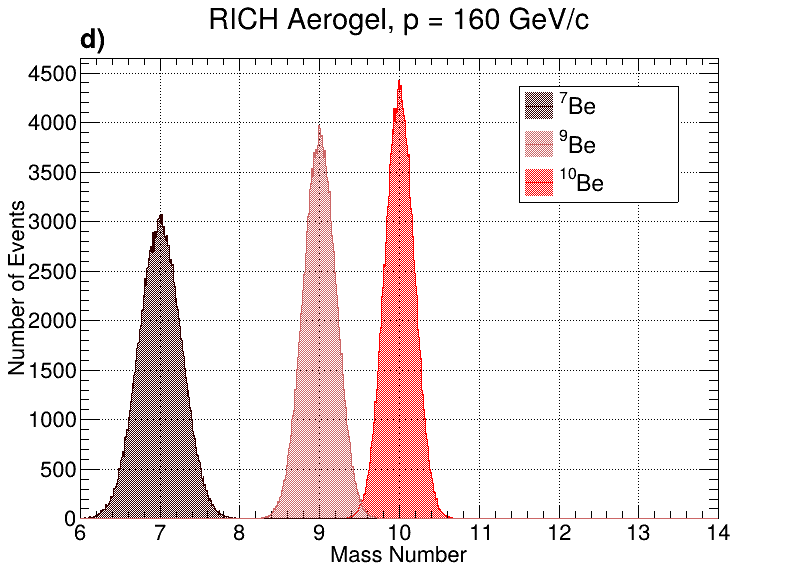}
    \caption{(\textbf{a}) Mass
 separation for beryllium isotopes as a function of the particle momentum. It compares the performance of three different detector systems: TOF (dotted lines), aerogel-RICH (solid lines) and NaF-RICH (dashed lines); (\textbf{b}--\textbf{d}) expected mass distribution for beryllium isotopes at specific particle momentum providing a more detailed look at the identification capability of the TOF at 10 GeV/c (\textbf{b}), NaF-RICH at 120 GeV/c (\textbf{c}) and aerogel-RICH at 160 GeV/c (\textbf{d}).}
    \label{fig:dM_M}
\end{figure}

\section{Conclusions}
\label{sec:conc}

We have demonstrated that the combination of a RICH detector with a TOF system is feasible using current SiPM technology and fast front-end electronics. While we presented a possible configuration for isotope identification, significant technical challenges remain. The performance figures are based on beam test data and a fast, simplified simulation. Although this preliminary work gives a hint that a TOF system with $\mathcal{O}(10~\text{ps})$ resolution could be built alongside a RICH detector with $\mathcal{O}(\text{mrad})$ angular resolution, further studies and optimizations are necessary to realize such a device. Specifically, these optimizations involve determining the overall detector size (e.g., configuring the TOF layer distance based on the detector time resolution) and the readout pixel size, with a critical focus on performing detailed performance studies with the spectrometer.

\section*{Author contributions}
Conceptualization, M.N.M., G.D.R., F.L. and N.N.; 
methodology, M.N.M., L.C., G.D.R., F.L. and N.N.;
software, L.C., G.D.R., M.G., A.L., L.L., N.N., G.P. and R.P.; 
validation, M.G., A.L., L.L., N.N., G.P. and R.P.;
formal analysis, M.G., A.L., L.L., N.N., G.P. and R.P.;
investigation, M.N.M., M.G., A.L., L.L., N.N., G.P. and R.P.; 
resources, M.N.M., M.G., A.L., L.L., N.N., G.P. and R.P.; 
data curation, M.N.M., M.G., A.L., L.L., N.N., G.P. and R.P.; 
writing---original draft preparation,  M.N.M.; 
writing---review and editing,  M.N.M.; 
visualization,  M.N.M. and N.N.; 
supervision, M.N.M.; 
project administration, M.N.M.; 
funding acquisition, M.N.M. 
All authors have read and agreed to the published version of the manuscript.

\section*{Funding}
The research leading to the beam test results received partial funding from the European Union’s Horizon Europe research and innovation programme under grant agreement No. 101057511.

\section*{Data availability}
The datasets generated during and/or analyzed during the current study are available from the corresponding author on reasonable request.

\section*{Acknowledgments}
The authors would like to thank the INFN Bari
staff for their contribution to the procurement and to the construction of the prototype. In particular, we thank D. Dell’Olio, M. Franco, N. Lacalamita, F. Maiorano, M. Mongelli, C. Pastore and R. Triggiani for their technical support. The authors gratefully acknowledge the ALICE 3 RICH group for the use of the beam test data.
The authors would like to thank Weeroc for contributing to the development of the Radioroc~2/picoTDC board and providing support for the operation of Radioroc~2.
They also thank  the CERN beam team group for providing the facilities and logistical support for the test.

\section*{Conflicts of interest}
The authors declare no conflicts of interest.

\section*{Abbreviations}
The following abbreviations are used in this manuscript:
\\

\noindent 
\begin{tabular}{@{}ll}
ADC & Analog to Digital Converter \\
DAQ & Data Acquisition \\
FEB & Front-End Board \\
HPK & Hamamatsu \\
ISS & International Space Station \\
LUT & Lookup Table \\
MPPC & Multi-Pixel Photon Counter \\
NaF & Sodium Fluoride \\
PDE & Photo Detection Efficiency \\
PID & Particle Identification \\
RA & Ring Array  \\
RICH & Ring Imaging CHerenkov \\
SiPM & Silicon PhotoMultiplier \\
TOF & Time of Flight \\
TOT & Time over Threshold
\end{tabular}

\bibliography{MazziottaASAPP2025.bib}

@misc{petiroc2a,
    title = "{Petiroc 2A by Omega and Weeroc, France}",
    howpublished = {\url{https://www.weeroc.com/read_out_chips/petiroc-2a/}}
}

@article{Fleury:2014hfa,
    author = "Fleury, J. and Callier, S. and de La Taille, C. and Seguin, N. and Thienpont, D. and Dulucq, F. and Ahmad, S. and Martin, G.",
    title = "{Petiroc and Citiroc: front-end ASICs for SiPM read-out and ToF applications}",
    doi = "10.1088/1748-0221/9/01/C01049",
    journal = "JINST",
    volume = "9",
    pages = "C01049",
    year = "2014"
}

@article{Saleem:2023pwt,
    author = "Saleem et al., T.",
    title = "{Study experimental time resolution limits of recent ASICs at Weeroc with different SiPMs and scintillators}",
    archivePrefix = "arXiv",
    primaryClass = "physics.ins-det",
    doi = "10.1088/1748-0221/18/10/P10005",
    journal = "JINST",
    volume = "18",
    number = "10",
    pages = "P10005",
    year = "2023"
}

@article{Altruda:2023qoh,
    author = "Altruda et al., S.",
    title = "{PicoTDC: a flexible 64 channel TDC with picosecond resolution}",
    doi = "10.1088/1748-0221/18/07/P07012",
    journal = "JINST",
    volume = "18",
    number = "07",
    pages = "P07012",
    year = "2023"
}

@article{DeRobertis:2018vls,
    author = "De Robertis et al., G.",
    editor = "Dalla Torre, S. and Gobbo, B. and Levorato, S. and Ropelewski, L. and Tessarotto, F.",
    title = "{A MOdular System for Acquisition, Interface and Control (MOSAIC) of detectors and their related electronics for high energy physics experiment}",
    doi = "10.1051/epjconf/201817407002",
    journal = "EPJ Web Conf.",
    volume = "174",
    pages = "07002",
    year = "2018"
}

@article{Altamura:2025ajm,
    author = "Altamura, A. R. and others",
    title = "{Beam test studies for a SiPM-based RICH detector prototype for the future ALICE~3 experiment}",
    doi = "10.1140/epjc/s10052-025-14287-7",
    journal = "Eur. Phys. J. C",
    volume = "85",
    number = "5",
    pages = "578",
    year = "2025"
}

@article{Pillera:2025iqx,
    author = "Pillera, R. and others",
    title = "{Beam test and performance assessment for the prototype of a novel compact RICH detector with timing capabilities for the future ALICE 3 PID system at HL-LHC}",
    doi = "10.1016/j.nima.2025.170708",
    journal = "Nucl. Instrum. Meth. A",
    volume = "1080",
    pages = "170708",
    year = "2025"
}

@article{Mazziotta:2025zxj,
    author = "Mazziotta, M. N. and others",
    title = "{Test beam performance of a novel RICH detector with timing capabilities for the future ALICE~3 PID system at~LHC}",
    doi = "10.1088/1748-0221/20/05/C05038",
    journal = "JINST",
    volume = "20",
    number = "05",
    pages = "C05038",
    year = "2025"
}

@article{Mazziotta:2025kdr,
    author = "Mazziotta, M. N. and others",
    title = "{Development of a novel compact and fast SiPM-based RICH detector for the future ALICE 3 PID system at LHC}",
    doi = "10.1088/1748-0221/20/01/C01001",
    journal = "JINST",
    volume = "20",
    number = "01",
    pages = "C01001",
    year = "2025"
}

@INPROCEEDINGS{Nicassio:2023kux,
  author={Nicassio, Nicola and Altamura, Anna Rita and Altomare, Corrado and Robertis, Giuseppe De and Bari, Domenico Di and Mauro, Antonello Di and Guerra-Pulido, Jaime Octavio and Mazziotta, Mario Nicola and Nappi, Eugenio and Paic, Guy and Pillera, Roberta and Volpe, Giacomo},
  booktitle={2023 9th International Workshop on Advances in Sensors and Interfaces (IWASI)}, 
  title={A combined SiPM-based TOF+RICH detector for future high-energy physics experiments}, 
  year={2023},
  volume={},
  number={},
  pages={199-204},
  doi={10.1109/IWASI58316.2023.10164558}
}

@article{ADACHI_Aerogel,
title = "{Status of high-quality silica aerogel radiators}",
journal = {NIM A},
volume = {952},
pages = {161919},
year = {2020},
issn = {0168-9002},
url = {https://doi.org/10.1016/j.nima.2019.02.046},
author = "{I. Adachi}",
}

@article{ALICE:2022wwr,
    collaboration = "ALICE",
    title = "{Letter of intent for ALICE 3: A next-generation heavy-ion experiment at the LHC}",
    eprint = "2211.02491",
    archivePrefix = "arXiv",
    primaryClass = "physics.ins-det",
    reportNumber = "CERN-LHCC-2022-009, LHCC-I-038",
    month = "11",
    year = "2022"
}

@article{Altamura:2024xtz,
    author = "Altamura, Anna Rita",
    title = "{Silica aerogel characterization for the ePIC dRICH detector}",
    doi = "10.22323/1.476.1116",
    journal = "PoS",
    volume = "ICHEP2024",
    pages = "1116",
    year = "2025"
}

@misc{ntc10k,
    title = "{TEWA TT4-10KC3-T125-M5-500 NTC Thermistor}",
    howpublished = "\url{https://www.tme.eu/Document/6d2e1b5322209ea0890793c756dbc659/TT4-10KC3-T125-M5-500.pdf}"
}

@misc{ads1115,
    title = "{ADS1115 16-Bit ADC - 4 Channel with Programmable Gain Amplifier - STEMMA QT / Qwiic}",
    howpublished = "\url{https://www.adafruit.com/product/1085}"
}

@misc{sht31,
    title = "{Adafruit Sensirion SHT31-D - Temperature \& Humidity Sensor}",
    howpublished = "\url{https://www.adafruit.com/product/2857}"
}

@misc{hlcd,
    title = "{Samtec HLCD-20-06.00-TR-TR-1 Cable}",
    howpublished = {\url{https://www.samtec.com/products/hlcd}}
}

@article{Mazziotta:2022vow,
    author = "Mazziotta, M. N. and others",
    title = "{A light tracker based on scintillating fibers with SiPM readout}",
    doi = "10.1016/j.nima.2022.167040",
    journal = "Nucl. Instrum. Meth. A",
    volume = "1039",
    pages = "167040",
    year = "2022"
}

@inproceedings{Pillera:2023hzp,
    author = "Pillera, Roberta and others",
    title = "{Characterization of a light fiber tracker prototype with SiPM array readout}",
    booktitle = "{9th International Workshop on Advances in Sensors and Interfaces}",
    doi = "10.1109/IWASI58316.2023.10164306",
    month = "6",
    year = "2023"
}

@article{Giovacchini:2023ixx,
    author = "Giovacchini, F.",
    collaboration = "AMS",
    title = "{The RICH detector of the AMS-02 experiment aboard the International Space Station}",
    doi = "10.1016/j.nima.2023.168434",
    journal = "Nucl. Instrum. Meth. A",
    volume = "1055",
    pages = "168434",
    year = "2023"
}

@article{Cossio:2024nju,
    author = "Cossio, F. and others",
    title = "{ALCOR: A mixed-signal ASIC for the dRICH detector of the ePIC experiment at the EIC}",
    doi = "10.1016/j.nima.2024.169817",
    journal = "Nucl. Instrum. Meth. A",
    volume = "1069",
    pages = "169817",
    year = "2024"
}

@article{delaTorreLuque:2022vhm,
    author = "de la Torre Luque, Pedro and Mazziotta, Mario Nicola and Ferrari, Alfredo and Loparco, Francesco and Sala, Paola and Serini, Davide",
    title = "{FLUKA cross sections for cosmic-ray interactions with the DRAGON2 code}",
    eprint = "2202.03559",
    archivePrefix = "arXiv",
    primaryClass = "astro-ph.HE",
    doi = "10.1088/1475-7516/2022/07/008",
    journal = "JCAP",
    volume = "07",
    number = "07",
    pages = "008",
    year = "2022"
}

@inproceedings{Credo:2004qgy,
    author = "Credo, Timothy and Frisch, Henry and Sanders, Harold and Schroll, Robert and Tang, Fukun",
    title = "{Picosecond time-of-flight measurement for colliders using Cherenkov light}",
    booktitle = "{2004 IEEE Nuclear Science Symposium and Medical Imaging Conference}",
    doi = "10.1109/NSSMIC.2004.1462263",
    number = "1",
    pages = "586--590",
    year = "2004"
}

@article{Carnesecchi:2023dfq,
    author = "Carnesecchi, F. and others",
    title = "{Measurements of the Cherenkov effect in direct detection of charged particles with SiPMs}",
    eprint = "2305.17762",
    archivePrefix = "arXiv",
    primaryClass = "physics.ins-det",
    doi = "10.1140/epjp/s13360-023-04397-0",
    journal = "Eur. Phys. J. Plus",
    volume = "138",
    number = "9",
    pages = "788",
    year = "2023"
}

@article{Gundacker:2020cnv,
    author = "Gundacker, Stefan and Heering, Arjan",
    title = "{The silicon-photomultiplier: fundamentals and applications of a modern solid-state photon detector}",
    doi = "10.1088/1361-6560/ab7b2d",
    journal = "Phys. Med. Biol.",
    volume = "65",
    number = "17",
    pages = "17TR01",
    year = "2020"
}

@article{AMS:2021nhj,
    author = "Aguilar, M. and others",
    collaboration = "AMS",
    title = "{The Alpha Magnetic Spectrometer (AMS) on the international space station: Part II {\textemdash} Results from the first seven years}",
    doi = "10.1016/j.physrep.2020.09.003",
    journal = "Phys. Rept.",
    volume = "894",
    pages = "1--116",
    year = "2021"
}

@article{Wakely:2023iaq,
    author = "Wakely, S. P. and others",
    title = "{Cosmic-ray Isotope Measurements with HELIX}",
    doi = "10.22323/1.444.0118",
    journal = "PoS",
    volume = "ICRC2023",
    pages = "118",
    year = "2023"
}

@article{Sadrozinski:2013nja,
    author = "Sadrozinski, H. F. -W. and others",
    editor = "Pace, Emanuele and Talamonti, Cinzia and Bruzzi, Mara",
    title = "{Ultra-fast silicon detectors}",
    doi = "10.1016/j.nima.2013.06.033",
    journal = "Nucl. Instrum. Meth. A",
    volume = "730",
    pages = "226--231",
    year = "2013"
}

@article{Pellegrini:2014lki,
    author = "Pellegrini, G. and others",
    editor = "Unno, Yoshinobu and Fukazawa, Yasushi and Hou, Suen and Ohsugi, Takashi and Sadrozinski, Hartmut F. -W.",
    title = "{Technology developments and first measurements of Low Gain Avalanche Detectors (LGAD) for high energy physics applications}",
    doi = "10.1016/j.nima.2014.06.008",
    journal = "Nucl. Instrum. Meth. A",
    volume = "765",
    pages = "12--16",
    year = "2014"
}

@article{vanDijk:2025ggb,
    author = "van Dijk, Maarten and Hayat, Ahsan and Banerjee, Dipanwita and Bernhard, Johannes and Gokturk, Berare and Nevay, Laurie and Petersen, Jorgen and Schwinzerl, Martin",
    title = "{Particle production and identification for the T10 secondary beamline of the CERN East Area}",
    eprint = "2507.02567",
    archivePrefix = "arXiv",
    primaryClass = "hep-ex",
    month = "7",
    year = "2025"
}

@article{Jeon:2023pie,
    author = "Jeon, H. B. and others",
    title = "{The Design and Status of the HELIX Ring Imaging Cherenkov Detector and Hodoscope Systems}",
    doi = "10.22323/1.444.0121",
    journal = "PoS",
    volume = "ICRC2023",
    pages = "121",
    year = "2023"
}

@article{Bindi:2010zzb,
    author = "Bindi, V. and others",
    title = "{The scintillator detector for the fast trigger and time-of-flight (TOF) measurement of the space experiment AMS-02}",
    doi = "10.1016/j.nima.2010.08.019",
    journal = "Nucl. Instrum. Meth. A",
    volume = "623",
    pages = "968--981",
    year = "2010"
}

@article{HELIX:2023twg,
    author = "Coutu, S. and others",
    collaboration = "HELIX",
    title = "{The High Energy Light Isotope eXperiment program of direct cosmic-ray studies}",
    eprint = "2312.06796",
    archivePrefix = "arXiv",
    primaryClass = "astro-ph.IM",
    doi = "10.1088/1748-0221/19/01/C01025",
    journal = "JINST",
    volume = "19",
    number = "01",
    pages = "C01025",
    year = "2024"
}

@article{Cerasole:2025tiz,
    author = "Cerasole, D. and others",
    title = "{Development of a light tracker based on thin scintillating fibers and Silicon Photomultipliers for space application}",
    doi = "10.1016/j.nima.2025.171064",
    journal = "Nucl. Instrum. Meth. A",
    volume = "1082",
    pages = "171064",
    year = "2026"
}

@article{Carnesecchi:2023jsu,
    author = "Carnesecchi, F. and others",
    title = "{A new low gain avalanche diode concept: the double-LGAD}",
    eprint = "2307.14320",
    archivePrefix = "arXiv",
    primaryClass = "physics.ins-det",
    doi = "10.1140/epjp/s13360-023-04621-x",
    journal = "Eur. Phys. J. Plus",
    volume = "138",
    number = "11",
    pages = "990",
    year = "2023"
}

@article{Strazzi:2024tmn,
    author = "Strazzi, Sofia",
    collaboration = "ALICE",
    title = "{Innovations in silicon detector technologies for next-generation experiments: Improving timing precision of LGADs for ALICE 3}",
    doi = "10.1016/j.nima.2024.169893",
    journal = "Nucl. Instrum. Meth. A",
    volume = "1069",
    pages = "169893",
    year = "2024"
}
\bibliographystyle{unsrt}

\end{document}